\documentclass[aps,prl,reprint,superscriptaddress,letterpaperm,showpacs]{revtex4-1}
\usepackage[english]{babel}
\usepackage{amsmath}
\usepackage{amsfonts}
\usepackage{amssymb}
\usepackage{graphicx}
\usepackage{bm}
\usepackage{bbm}
\usepackage{empheq}

\begin{document}

\title{Photon Emission from a Cavity-Coupled Double Quantum Dot}

\author{Y.-Y. Liu}
\affiliation{Department of Physics, Princeton University, Princeton, New Jersey 08544, USA}
\author{K. D. Petersson}
\altaffiliation{Present address: Center for Quantum Devices, Niels Bohr Institute, University of Copenhagen, Blegdamsvej 17, DK-2100 Copenhagen $\O$, Denmark}
\affiliation{Department of Physics, Princeton University, Princeton, New Jersey 08544, USA}
\author{J. Stehlik}
\affiliation{Department of Physics, Princeton University, Princeton, New Jersey 08544, USA}
\author{J. M. Taylor}
\affiliation{Joint Quantum Institute/NIST, College Park, Maryland 20742, USA}
\author{J. R. Petta}
\affiliation{Department of Physics, Princeton University, Princeton, New Jersey 08544, USA}

\date{\today}

\begin{abstract}
We study a voltage biased InAs double quantum dot (DQD) that is coupled to a superconducting transmission line resonator. Inelastic tunneling in the DQD is mediated by electron phonon coupling and coupling to the cavity mode. We show that electronic transport through the DQD leads to photon emission from the cavity at a rate of 10 MHz. With a small cavity drive field, we observe a gain of up to 15 in the cavity transmission. Our results are analyzed in the context of existing theoretical models and suggest that it may be necessary to account for inelastic tunneling processes that proceed via simultaneous emission of a phonon and a photon.
\end{abstract}


\maketitle


Cavity quantum electrodynamics (cavity-QED) explores quantum optics at the most basic level of a single photon interacting with a single atom \cite{Walls1995}. In a conventional laser, population inversion of a large ensemble of atoms provides an optical gain medium via stimulated emission. With just one or two atoms in a laser cavity, novel quantum optical effects can be observed. In cavity-QED, lasing has been achieved for single Rydberg atoms passing through a superconducting microwave cavity \cite{Meschede1985}, and a single Ca atom strongly coupled to a high finesse optical cavity \cite{McKeever2003}. Thresholdless lasing and antibunching of the emitted photons were observed in the single atom limit. Lasing has also been achieved in a solid-state device using a quantum dot emitter as a single `artificial atom' inside a micropillar cavity \cite{Ates2009}.

Circuit quantum electrodynamics (cQED) exploits high quality factor superconducting resonators to realize strong coupling between microwave photons and a solid-state quantum device \cite{Blais2004}. For superconducting qubits, the resonator enables qubit state readout and non-local qubit entanglement \cite{Wallraff2004, DiCarlo2009}. In turn, superconducting qubits can be used to program non-classical photon states within the cavity \cite{Houck2007, Hofheinz2009, Bozyigit2011}. Experiments have also used microwave cavities to explore photon emission from voltage-biased superconducting circuits \cite{Astafiev2007,Hofheinz2011}. More recently, quantum dots have been integrated with superconducting microwave cavities, with large charge-cavity couplings $g_0/2\pi$ $\sim 20-100$ MHz \cite{Frey2012, Toida2013, Petersson2012, Viennot2013, Deng2013}. Spin-state readout \cite{Petersson2012} and non-local coupling of distant quantum dot circuits \cite{Delbecq2013} have also been demonstrated.

\begin{figure}[t]
  \begin{center}
		\includegraphics[width=\columnwidth]{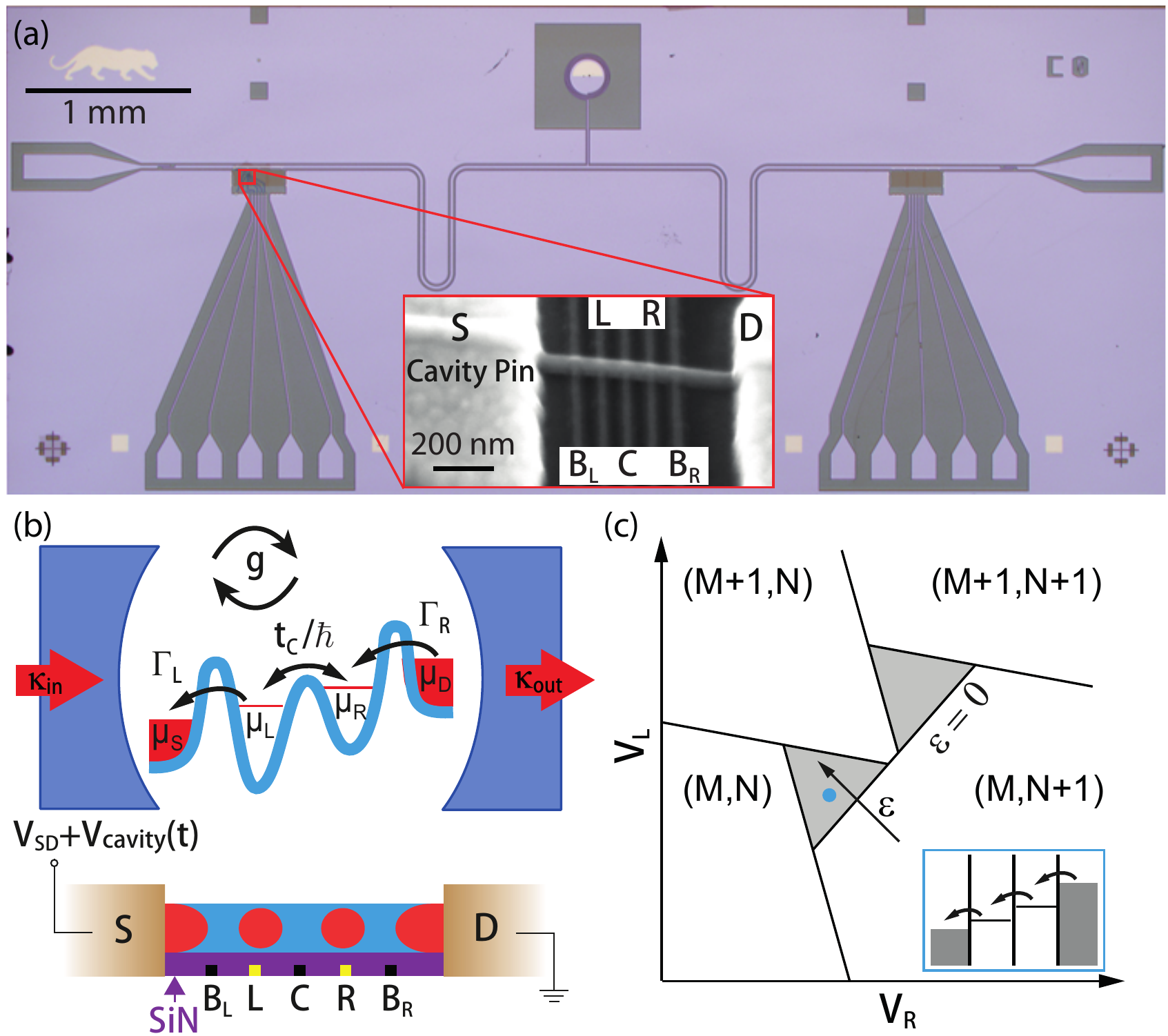}
  \caption{\label{Fig: scheme} (a) Optical micrograph of the hybrid system. Inset: Scanning electron micrograph of an InAs nanowire DQD. (b) The DQD is formed by biasing gates $B_{\rm L}$, $B_{\rm R}$ and  $C$ at negative voltages to form left and right tunnel barriers with rates $\Gamma_{\rm L}$ and $\Gamma_{\rm R}$, and an interdot tunnel barrier with rate \textbf{$t_{\rm c}/\hbar$}. Cavity photons are coupled to the input and output ports with rates $\kappa_{\rm in}$ and $\kappa_{\rm out}$. A source-drain bias $V_{\rm SD} = (\mu_{\rm D}-\mu_{\rm S})/e$ is applied to the device. (c) Schematic of the charge stability diagram near the ($M$, $N$+1)$\leftrightarrow$($M$+1, $N$) interdot charge transition. Sequential tunneling is allowed within FBTs in the charge stability diagram (grey triangles). Inset: DQD energy level configuration in the lower FBT.}
  \end{center}
\end{figure}

It is well known that electron tunneling in semiconductor DQDs can be driven by the absorption of microwave photons in a process called photon assisted tunneling (PAT) \cite{Kouwenhoven1994, Blick1995, Wiel2002}. In this Letter, we investigate the inverse process, and show that dc transport of electrons leads to photon emission in a cavity-coupled InAs nanowire DQD. Previous work on semiconductor DQDs showed that inelastic interdot tunneling processes are mediated by spontaneous emission of a phonon \cite{Brandes1999, Fujisawa1998}. DQDs have also been used as frequency-selective single photon detectors \cite{Gustavsson2007}. In our system, a charge-cavity coupling rate $g_0/2\pi$ $\sim$ 16 MHz opens up an additional channel for dissipation. Remarkably, we observe a gain of up to 15 in cavity transmission near the interdot charge transition with a device current $I$ $\sim$ 8 nA \cite{SOM}. Additionally, in the absence of a cavity drive, we directly measure a photon emission rate $\Delta\Gamma_{\rm p}$ $\sim$ 10 MHz. Our experimental results show that the cavity-coupled DQD provides fertile ground for exploring quantum optics in condensed matter systems, such as nonclassical states of light \cite{Nori2011}.

The hybrid device is shown in Fig.\ \ref{Fig: scheme}(a). A half-wavelength superconducting Nb transmission line resonator has a center frequency $f_c$ = 7862 MHz and quality factor Q $\sim$ 3600. Five Ti/Au bottom gates ($B_{\rm L}$, $L$, $C$, $R$, $B_{\rm R}$) selectively deplete an InAs nanowire resulting in a double well confinement potential, as shown in Fig.\ \ref{Fig: scheme}(b) \cite{Fasth2007, Nadj-Perge2010}. An excess charge trapped in the DQD interacts with the electric field of the resonator leading to a large charge-cavity coupling rate $g_0/2\pi\sim 16$ MHz \cite{Petersson2012, SOM}. The device is measured in a dilution refrigerator with a base temperature of 10 mK.

\begin{figure}[t]
  \begin{center}
		\includegraphics[width=\columnwidth]{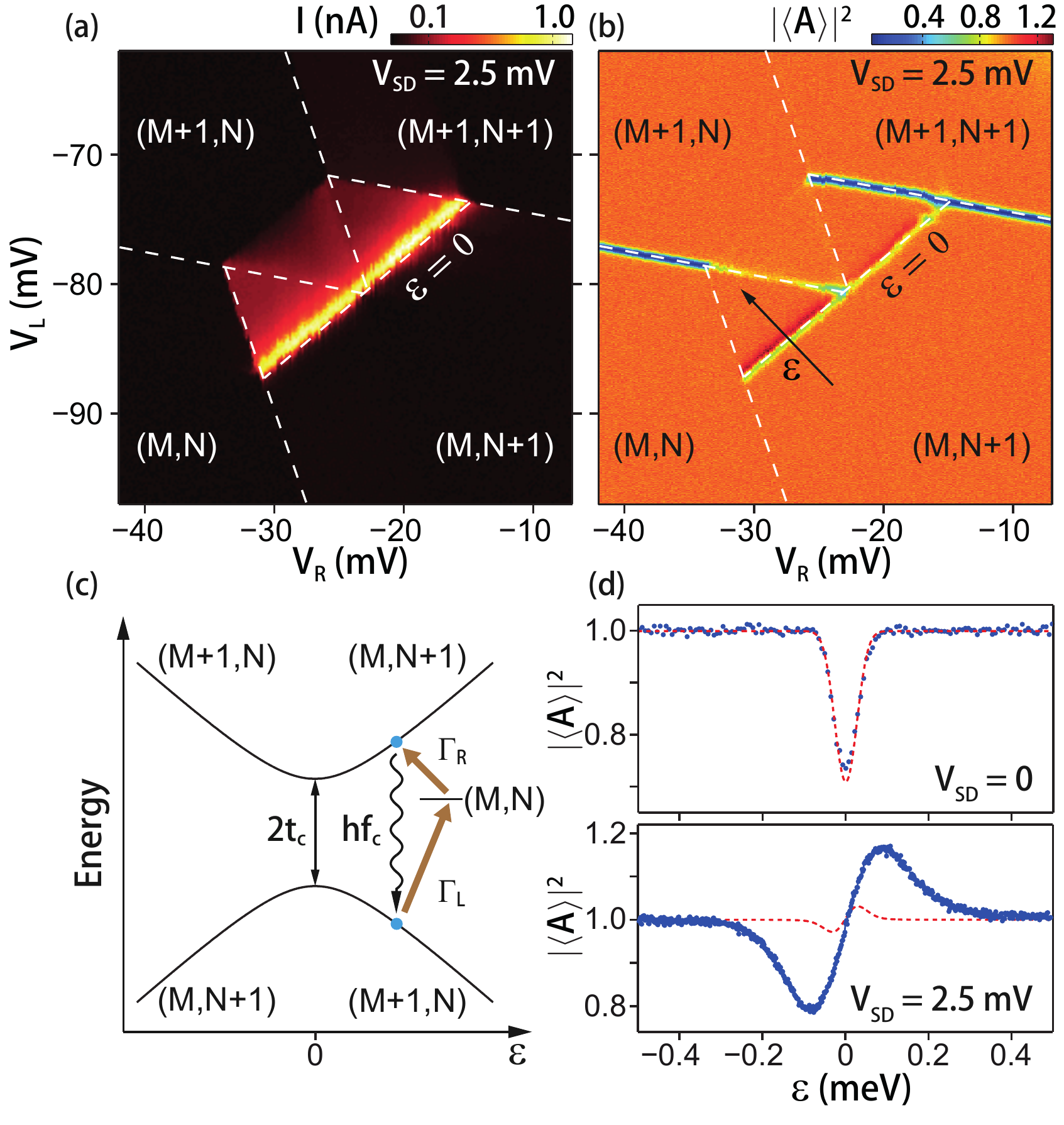}
  \caption{\label{Fig: well defined area} (a) DQD current $I$ plotted as a function of $V_{\rm L}$ and $V_{\rm R}$ with $V_{\rm SD}$ = 2.5 mV. (b) The corresponding normalized transmission $|\langle A\rangle|^2$. Gain is observed at positive detuning ($\epsilon > 0$). (c) Energy level diagram at the ($M$, $N$+1)$\leftrightarrow$($M$+1, $N$) interdot charge transition, illustrating a possible gain mechanism. (d) $|\langle A\rangle|^2$ as a function of $\epsilon$ with $V_{\rm SD}$ = 0 (upper panel) and $V_{\rm SD}$ = 2.5 mV (lower panel). Dashed lines are best fits to theory (see main text). At finite bias, the model underestimates the gain and range of detuning where gain occurs by a factor of $\sim$ 4.}
  \end{center}
\end{figure}

With a source-drain bias applied across the DQD, sequential tunneling is allowed within finite bias triangles (FBT) in the charge stability diagram, as schematically shown in Fig.\ 1(c) \cite{Wiel2002}. Our experiments are performed in the many-electron regime and we label the DQD charge states $(N_{\rm L}, N_{\rm R})$, where $N_{\rm L}$($N_{\rm R}$) denote the number of electrons in the left(right) quantum dot. In the lower FBT, DQD transport follows the cycle ($M$, $N$) $\rightarrow$ ($M$, $N$+1) $\rightarrow$ ($M$+1, $N$) $\rightarrow$ ($M$, $N$). In Fig.\ \ref{Fig: well defined area}(a) we plot $I$ as a function of $V_{\rm L}$ and $V_{\rm R}$  with $V_{\rm SD}$ = 2.5 mV, revealing the FBTs. Well outside of the FBTs, where $I$ = 0, the DQD is deep in Coulomb blockade and the charge state is fixed. Large tunnel couplings to the leads results in some cotunneling current between the FBTs.

Electronic transport can be driven by the absorption of a photon in PAT \cite{Kouwenhoven1994,Blick1995}. Here we measure cavity transmission in the presence of a source-drain bias to determine if electronic transport results in photon emission. We apply a microwave drive at $f_c$ = 7862 MHz with a power $P \approx -105$ dBm and measure the amplitude $A$ of the transmitted field using heterodyne detection \cite{SOM}. Cavity transmission $|\langle A\rangle|^2$ is plotted as a function of $V_{\rm L}$ and $V_{\rm R}$ in Fig.\ 2(b). The cavity transmission is normalized relative to the value measured deep in Coulomb blockade, where the DQD is effectively decoupled from the cavity \cite{Petersson2012}. Cavity transmission is reduced at charge transitions that change the total electron number, consistent with previous work \cite{Frey2012, Toida2013, Petersson2012}. However, in contrast with previous work, we observe gain $|\langle A\rangle|^2>1$ along the positive detuning side of the interdot charge transition where electron transport proceeds downhill in energy ($\epsilon$ and $V_{\rm SD}$ $>$ 0), indicating that photon emission is related to the ($M$+1, $N$) $\leftrightarrow$ ($M$, $N$+1) interdot charge transition.

The detuning dependence (in lever-arm corrected units of meV) of the cavity transmission is investigated in Fig.\ 2(d) for $V_{\rm SD}$ = 0 (upper panel) and $V_{\rm SD}$ = 2.5 mV (lower panel) \cite{SOM}. For $V_{\rm SD}$ = 0, $|\langle A\rangle|^2$ is reduced near $\epsilon$ = 0, consistent with previous work \cite{Petersson2012}. In comparison, the data acquired with $V_{\rm SD}$ = 2.5 mV show gain $|\langle A\rangle|^2$ $>$ 1 for $\epsilon$ $>$ 0 and a damping $|\langle A\rangle|^2$ $<$ 1  for $\epsilon$ $<$ 0. The gain that is observed for $\epsilon$ $>$ 0 indicates that the DQD is transferring energy to the cavity mode during the downhill inelastic interdot tunneling process.

Qualitatively, the DQD can be modeled as a charge qubit with Hamiltonian $H$ = $\frac{\epsilon}{2} \sigma_z$ + $t_{\rm c}\sigma_x$, where $\sigma_{x}$ and $\sigma_{z}$ are the Pauli matrices. This Hamiltonian results in a detuning dependent energy splitting ${\Omega(\epsilon)=\sqrt{\epsilon^2+4t_{\rm c}^2}}$. From conservation of energy, we anticipate strong emission into the cavity when $hf_c = \Omega \approx$ 33 $\mu$eV. With $t_{\rm c}$ = 16.4 $\mu$eV, this corresponds to $\epsilon\sim1$ $\mu$eV. Near $\epsilon$ = 0 $\mu$eV elastic tunneling processes dominate \cite{Fujisawa1998}, while at far detuning the effective charge-cavity interaction rate $g$ = $g_0\frac{2t_{\rm c}}{\Omega}$ vanishes \cite{Childress2004, Jin2011, Petersson2012,  Schroer2012}. We therefore expect photon gain effects to be the strongest for  $0 \lesssim \epsilon \lesssim 30$ $\mu$eV. However, we observe a peak in transmission at $\epsilon$ $\sim$ 80 $\mu$eV. Similarly, photon absorption should be the strongest for $-30 \lesssim \epsilon \lesssim 0$ $\mu$eV.  Surprisingly, microwave amplification and absorption both extend over a $\sim$ 200 $\mu$eV range of detuning.

We model the zero-bias transmission data using the Jaynes-Cummings Hamiltonian with an effective charge-cavity interaction rate $g$ = $g_0\frac{2t_{\rm c}}{\Omega}$ \cite{Childress2004, Jin2011, Petersson2012,  Schroer2012}. The model assumes a phonon relaxation rate $\gamma/2\pi \approx70$ MHz \cite{Petta2004} and coupling to a single dominant resonator mode of frequency $f_c$ = 7862 MHz with a total decay rate \newline $\kappa/2\pi = f_c/Q = (\kappa_{\rm in}+\kappa_{\rm out}+\kappa_{\rm i})/2\pi \approx 2$ MHz. We assume the cavity is symmetric with $\kappa_{\rm in}=\kappa_{\rm out}$, and neglect internal loss ($\kappa_i=0$). Low frequency charge noise is accounted for by smoothing the fit function using a Gaussian with standard deviation $\sigma_{\epsilon}$ = 25 $\mu$eV \cite{Petersson2012}. As shown in Fig.\ 2(d), the model is in excellent agreement with the zero-bias data, yielding best fit values of $t_{\rm c}=16.4$ $\mu$eV and $g_0/2\pi=16$ MHz.

To model the sequential tunneling dynamics in the lower FBT, we consider a transport process proposed by Jin \textit{et al.} that ``repumps" the DQD into the excited state ($M$, $N$+1) \cite{Childress2004,Jin2011,Vavilov,SOM}. A complete transport cycle is shown in the level diagram in Fig.\ \ref{Fig: well defined area}(c). The DQD is pumped via a two-step incoherent tunneling process ($M$+1, $N$) $\rightarrow$ ($M$, $N$) $\rightarrow$ ($M$, $N$+1) with rates $\Gamma_{\rm L}$ and $\Gamma_{\rm R}$, respectively. At far detuning, these two processes are equivalent to pumping from ($M$+1, $N$) to ($M$, $N$+1) with an effective pump rate $\Gamma_{\rm eff}=\Gamma_{\rm R}\Gamma_{\rm L}/(\Gamma_{\rm R}+\Gamma_{\rm L})$ if the dwell time in ($M$, $N$) is short enough to be neglected. In the absence of other decay mechanisms, such as phonons, the electron tunnels from ($M$, $N$+1) to ($M$+1, $N$) by emitting a photon into the cavity mode to complete the transport cycle.

In terms of electron transport, our system is similar to the voltage biased Cooper pair box, where the voltage bias generates population inversion, producing a lasing state within the cavity \cite{Astafiev2007}. Population inversion is achieved through a cycle that changes the relative number of Cooper pairs on the island by 1. Enhanced photon emission observed in cavity-coupled Josephson junctions has also been associated with Cooper pair tunneling events \cite{Hofheinz2011} and recently investigated theoretically \cite{Armour2013,Gramich2013}. While there are similarities between the superconducting and semiconductor systems, the electron-phonon interaction is known to strongly influence charge and spin dynamics in semiconductor DQDs leading to complex behavior \cite{Brandes1999, Fujisawa1998, Wiel2002, Barrett2006, Colless2013}.

Predictions from the three-level model for $V_{\rm SD}$ = 2.5 mV are shown in the lower panel of Fig.\ \ref{Fig: well defined area}(d) \cite{SOM}. Values of $\kappa$, $\gamma$, $t_{\rm c}$ and $g_0$ are constrained by the $V_{\rm SD}$ = 0 data set. As before, the amplitude response function is smoothed using a Gaussian with width $\sigma_{\epsilon}$ = 25 $\mu$eV to account for charge noise. Taking $\Gamma_{\rm L}=\Gamma_{\rm R}$ as the only free parameters, the best fit has $\Gamma_{\rm L}/2\pi = \Gamma_{\rm R} /2\pi= 4$ GHz, in agreement with the values determined from the dc transport data \cite{SOM}. The experimental data have gain that is 4 times larger than theory and the range of detuning over which gain is observed is 3 -- 4 times broader in the experiments as well. Increasing the charge noise broadens the gain feature, but also reduces the level of gain, and is unable to account for the discrepancy.

The strong amplification and the broad linewidth in $\epsilon$ suggest that when the energy splitting of the DQD is 3--4 times the cavity energy, the system is still emitting photons effectively. Two potential contributions for this broadening include phonon-assisted and photon-assisted tunneling processes. Given previous work by Fujisawa \textit{et al.} and Petta \textit{et al.}, it is known that phonon emission leads to charge relaxation rates on the order of 100 MHz \cite{Fujisawa1998, Petta2004}. The charge relaxation rate is $\sim$ 6 times faster than $g_0/2\pi$ = 16 MHz and competes with the charge-cavity coupling rate. One natural interpretation is that the DQD relaxes through emission of a phonon and a photon \cite{Barrett2006, Colless2013}. A wider range of $\epsilon$ is then permitted for the resonant emission of photons. It is unlikely that the broad linewidth is due to shot noise, since we observe photon emission in specific hot spots in the charge stability diagram [see Fig.\ 3(a)], but measure a large current over a much broader detuning range \cite{Zakka2007}.

\begin{figure}[t]
  \begin{center}
		\includegraphics[width=\columnwidth]{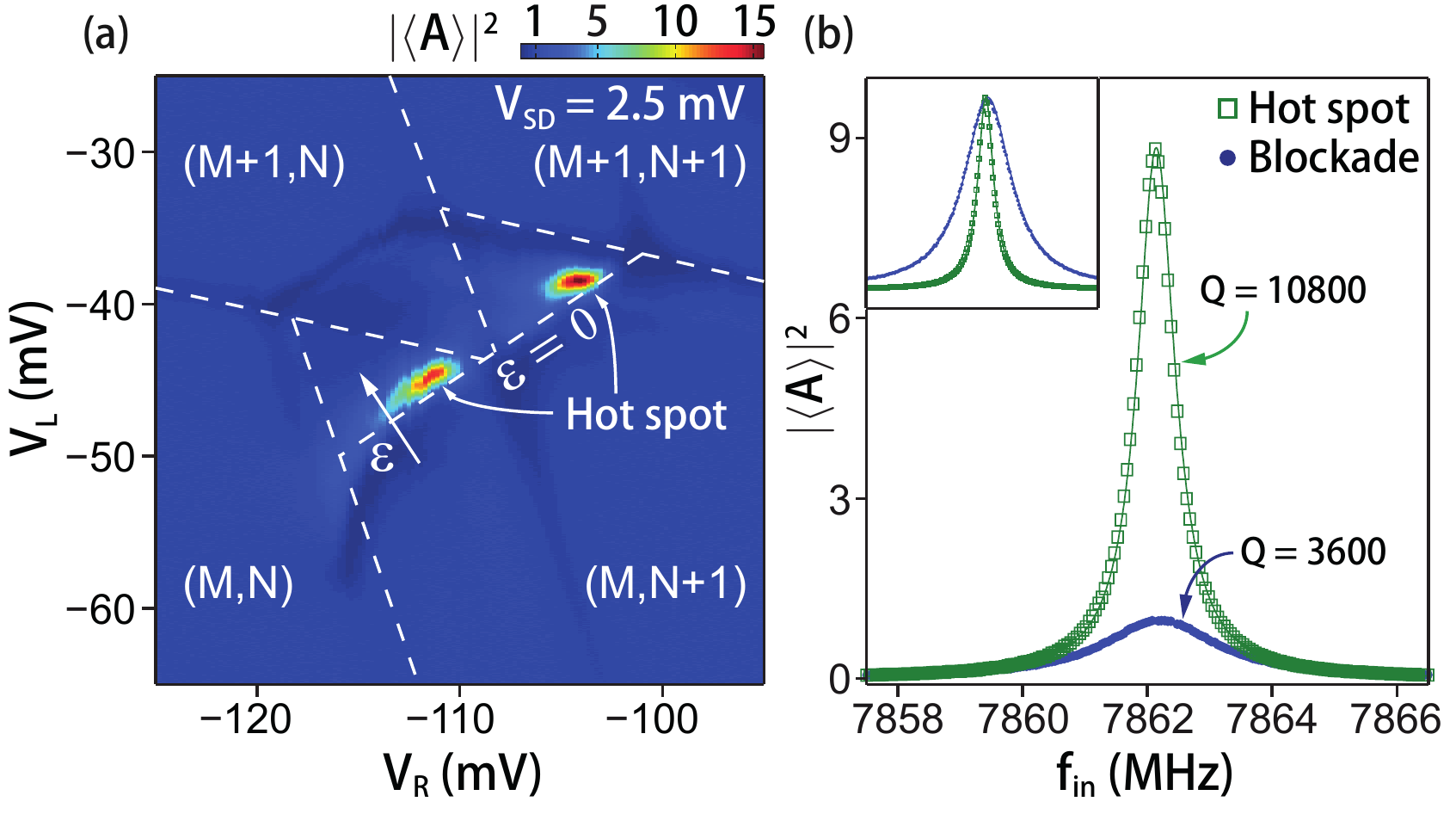}
  \caption{\label{Fig: highgain} (a) $|\langle A\rangle|^2$ as a function of $V_{\rm L}$ and $V_{\rm R}$ with the device configured to have larger tunnel rates to the leads. `Hot spots' with gain above 15 are observed. (b) $|\langle A\rangle|^2$ measured as a function of $f_{\rm in}$ near a hot spot (squares) and deep in Coulomb blockade (circles). Solid lines are fits to a Lorentzian. Inset: The same data are plotted with normalized peak amplitudes. At the hot spot, the cavity linewidth is reduced by a factor of $\sim$ 3.}
  \end{center}	
\end{figure}

To achieve higher photon emission rates we increased the tunnel coupling to the leads, resulting in an increase of the current through the DQD to $\sim$8 nA. Figure \ref{Fig: highgain}(a) shows $|\langle A\rangle|^2$ as a function of $V_{\rm L}$ and $V_{\rm R}$ in this more strongly tunnel coupled regime. At positive detuning, we observe a cavity gain of up to 15 at the cavity center frequency near `hot spots' in the charge stability diagram. We further investigate the gain mechanism by measuring $|\langle A\rangle|^2$ as a function of the drive frequency $f_{\rm in}$, as shown in Fig.\ \ref{Fig: highgain}(b). The blue curve (circular data points) has a peak transmission $|\langle A_0^{\rm CB}\rangle|^2 = 1$ with the dot biased deep in Coulomb blockade, yielding $Q_{\rm CB}$ = 3600. In contrast, the green curve (square data points) is acquired near the upper right `hot spot' in Fig.\ \ref{Fig: highgain}(a). Here, the peak transmission $|\langle A_0^{\rm HS}\rangle|^2$ $\sim$ 9 and the transmission peak is narrower, yielding $Q_{\rm HS}$ = 10,800. The narrower linewidth can be understood by the fact that $Q$ is the ratio of the average energy in the cavity to the cavity decay rate. Photon emission from the DQD increases the energy stored in the cavity and therefore the quality factor. Low-loss cavity theory predicts a gain-bandwidth product that is fixed by the cavity decay rate according to the expression $|\langle A_0\rangle| (2 \pi f_c / Q) =\kappa$ or $|\langle A_0^{\rm HS}\rangle|/|\langle A_0^{\rm CB}\rangle|= Q_{\rm HS}/Q_{\rm CB}$ \cite{Siegman1986}. For these device settings we measure $Q_{\rm HS}$/$Q_{\rm CB}$ = 3 = $|\langle A_0^{\rm HS}\rangle|/|\langle A_0^{\rm CB}\rangle|$, in excellent agreement with theory.

\begin{figure}[t]
  \begin{center}
		\includegraphics[width=\columnwidth]{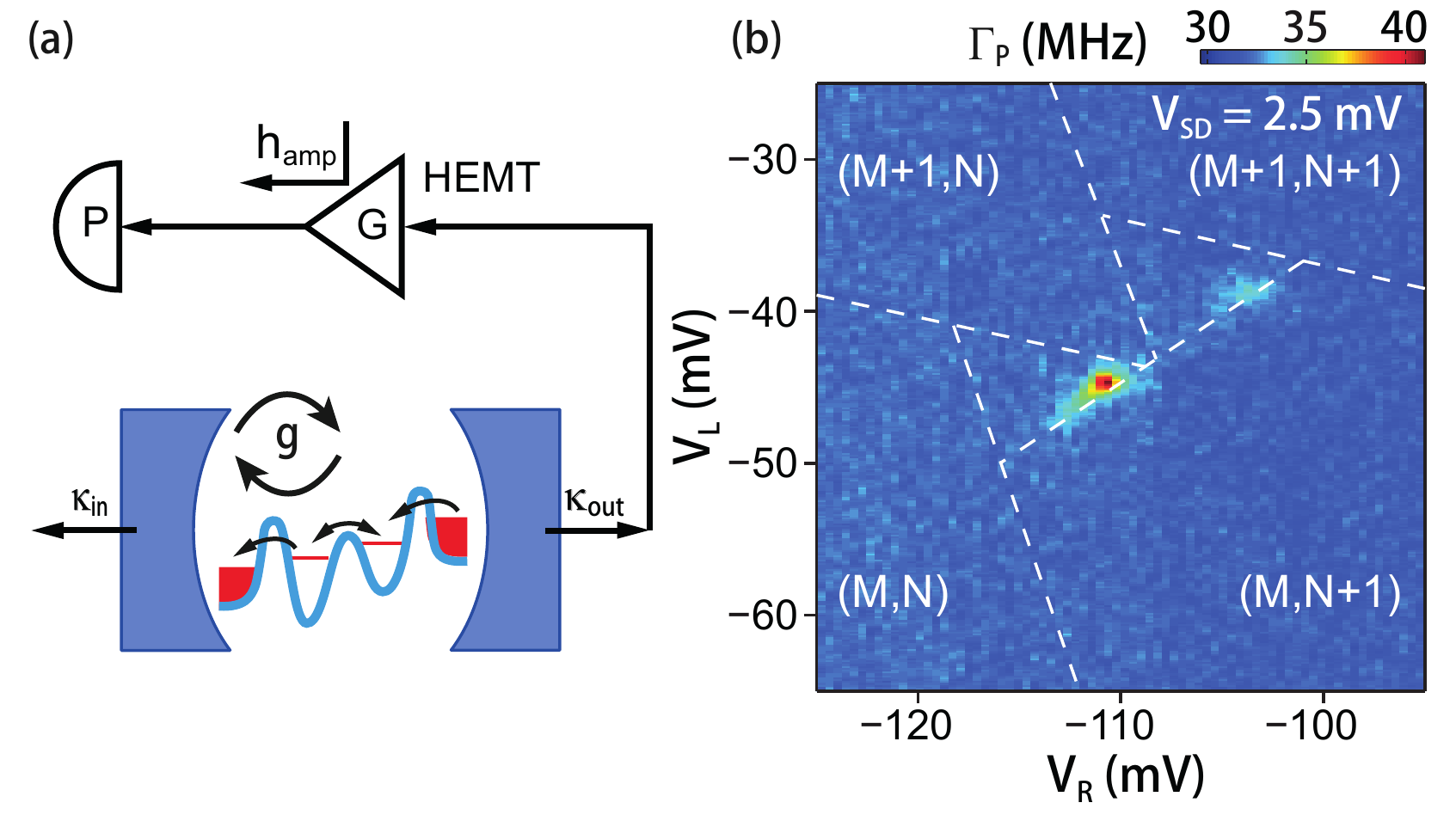}
  \caption{\label{Fig: emission} Photon emission from the voltage biased DQD in the absence of a cavity drive. (a) Simplified schematic of the experimental setup. Photons are emitted from the input and output ports of the cavity. Photons from the output port are amplified with gain $G$ (adding noise $h_{\rm amp}$) and then detected by a spectrum analyzer. (b) Photon emission rate plotted as a function of $V_{\rm L}$ and $V_{\rm R}$ for the same device configuration as in Fig.\ \ref{Fig: highgain}. The photon emission rate exceeds the background noise floor of the HEMT amplifier by $\sim$ 10 MHz at the hot spots.}
  \end{center}
\end{figure}

Given the large gains observed in experiment, we searched for direct evidence of photon emission from the DQD in the absence of a cavity drive tone. The measurement setup is shown in Fig.\ \ref{Fig: emission}(a). The output port of the cavity is connected to a high electron mobility transistor (HEMT) amplifier and the resulting signal is detected using a microwave spectrum analyzer. The photon emission rate is plotted as a function $V_{\rm L}$ and $V_{\rm R}$ in Fig.\ \ref{Fig: emission}(b). At the `hot spot', we measure a photon emission rate $\Delta\Gamma_{\rm p} \approx 10$ MHz above the background noise floor of the cryogenic HEMT amplifier (noise temperature $T_{\rm N}$ = 4 K). For a total cavity decay rate $\kappa/2\pi \approx 2$ MHz, the estimated photon number inside of the cavity is $N_{\rm p}$ = $2\Delta\Gamma_{\rm p}/\kappa\sim$ 2 assuming the cavity is symmetric. We take this estimate of the photon number as a lower bound since it does not account for the internal loss of the cavity and the line losses between the device and HEMT. We also note that $N_{\rm p}$ is much higher than the thermal occupation number $1/[\exp(hf_c/k_{\rm B}T)-1]$ $\ll$ 1.

Our data strongly suggest that inelastic tunneling in the cavity-coupled DQD results in photon emission. The photon emission efficiency $\beta$ can be estimated from the ratio of the photon emission rate to the electronic transport rate. The electron current $I$ at the `hot spot' is $\sim 8$ nA and thus $\beta\geq 2\Delta\Gamma_{\rm p}/(I/e)$ $\sim$ $0.4\times 10^{-3}$. This result should be contrasted with the Cooper pair box system, where photon emission is dominant and the efficiency is much closer to unity $\beta>0.4$ \cite{Astafiev2007}. The low efficiency of our cavity-coupled DQD suggests that other decay channels are stronger than photon emission into the cavity.

Previous theoretical work suggests that it may be possible to make a DQD laser with our device structure \cite{Childress2004,Jin2011}. We use a standard laser rate-equation model to determine if the cavity-coupled DQD is below the stimulated emission threshold \cite{Siegman1986}. The total photon emission rate from the DQD is $\Gamma_{\rm tot} = \Gamma_{\rm spon} + \Gamma_{\rm stim}$, where $\Gamma_{\rm spon}$ is the spontaneous emission rate (from the excited state to the ground state) and $\Gamma_{\rm stim} = \Gamma_{\rm spon} N_{\rm p}$ is the stimulated emission rate. The normalized inversion ratio is defined by $r\equiv\Gamma_{\rm spon}/\kappa$. When $r<1$ the system is below the stimulated emission threshold and $|\langle A_0 \rangle|= \kappa/(\kappa-\Gamma_{\rm spon})= 1/(1-r)$ \cite{Siegman1986}. Near the hot spot shown in Fig.\ \ref{Fig: highgain} $|\langle A_0\rangle|^2$ ranges from 9 to 15. From theory, this yields a normalized inversion ratio $r\approx2/3\sim3/4$ and $\Gamma_{\rm spon}/2\pi\approx1\sim2$ MHz. We estimate $N_{\rm p}$ = 2 from the data shown in Fig.\ \ref{Fig: emission}(b), which is in good agreement with the theoretical prediction $N_{\rm p}=r/(1-r)\approx2\sim3$ \cite{Siegman1986}. Lastly, we can use Fermi's golden rule to estimate $\Gamma_{\rm spon}$. Given the charge-cavity interaction rate $g_0/2\pi \approx 25$ MHz and a DQD decay rate $\gamma/2\pi\approx$ $0.1\sim1$ GHz \cite{Hayashi2003}, the predicted spontaneous emission rate (from the excited state to the ground state) is $\Gamma_{\rm spon}/2 \pi =4g^2/2 \pi \gamma \approx 2$ MHz, consistent with the estimate above. These simple estimates all suggest that the cavity-coupled DQD is below the stimulated emission threshold ($r$ $<$ 1).

In summary, we have investigated interactions between the dipole moment of a single excess electron in a DQD and the electromagnetic field of a microwave cavity. We observe a gain as large as 15 in the cavity transmission and also directly observe photon emission with a rate of 10 MHz above the noise floor of the HEMT amplifier. The gain observed in the cavity transmission is correlated with the interdot tunneling process, suggesting that inelastic current flow can proceed via emission of a photon or a phonon. Future experiments will explore the emission spectrum \cite{Astafiev2007} and quantum statistics of the output photon field \cite{McKeever2003, Jin2011, Bozyigit2011, Eichler2011}. Through further improvements in the photon emission efficiency, it may be possible to realize microwave amplifiers or on-demand single photon sources through single electron pumping in the DQD.
\raggedbottom

\begin{acknowledgments}
Research at Princeton was supported by the Sloan and Packard Foundations, Army Research Office Grant No.\ W911NF-08-1-0189, DARPA QuEST Grant No.\ HR0011-09-1-0007, and the NSF through DMR-0819860 and DMR-0846341. Partially sponsored by the United States Department of Defense. The views and conclusions contained in this document are those of the authors and should not be interpreted as representing the official policies, either expressly or implied, of the U.S. Government.
\end{acknowledgments}

\end{document}